\documentclass[10pt,a4paper]{article}

\usepackage{fullpage}
\usepackage{amsmath}
\usepackage{amssymb}
\usepackage{graphicx}
\usepackage[all,poly,curve,knot,cmtip]{xy}
\usepackage{citesort}

\newcommand {\be} {\begin{eqnarray*}}
\newcommand {\ee} {\end{eqnarray*}}
\newcommand {\bea} {\begin{eqnarray}}
\newcommand {\eea} {\end{eqnarray}}

\newcommand{\bm}[1]{\boldsymbol{#1}}

\newcommand{\pdiff}[2]{\frac{\partial{#1}}{\partial{#2}}}

\title{Induced current and redefinition of electric and magnetic fields from non-compact
Kaluza-Klein theory: An experimental signature of the fifth dimension}

\author{\textbf{Tom$\acute{\mbox{a}}\check{\mbox{s}}$ Liko}${}^{\dagger}$\\ \\
{\small \it Department of Physics, University of Waterloo}\\
{\small \it Waterloo, Ontario, Canada N2L 3G1}\\
{\small \it ${}^{\dagger}$Email correspondence: tliko@uwaterloo.ca}}

\begin{document}

\maketitle

\begin{abstract}

The field equations for gravitation and electromagnetism with sources in four
dimensions can be interpreted as arising from the vacuum Einstein equations in
five dimensions.  Gauge invariance of the electromagnetic potentials leads to
a ``generalized'' electromagnetic field tensor.  We use the action principle to
derive the equations of motion for free electromagnetic fields in flat spacetime,
and isolate an effective electromagnetic current with a source that is purely
higher-dimensional in origin.  This current provides, at least in principle,
a means of detecting extra dimensions experimentally.

\end{abstract}

\hspace{0.3cm}\textbf{PACS}: 04.50.$+$h; 11.10.Ef; 03.50.De

\hspace{0.3cm}\textbf{Keywords}: Induced-matter theory; Free electromagnetic
fields; Photon propagator

\section{Introduction}

A fashionable idea in today's modern theories of unification is that the world in which
we live is intrinsically higher-dimensional.  Prime among these are the superstring
theories \cite{hatfield,polchinski}, which require ten dimensions for mathematical
consistency.  Here six of the ten dimensions are compactified in circles of tiny radii
(on the order of the Planck length), thus leaving the four dimensions of space and time
that we experience.  Currently under investigation are unified theories which do not
assume the extra unobserved dimensions to be compactified.  One such approach with a
large extra dimension is Space-Time-Matter theory.  This was proposed by Wesson
\cite{wesson}.  The name stems from the geometrical interpretation of matter as an
induced quantity derived from the fifth dimension.  This point of view introduces a
certain symmetry in physics, since in mechanics we normally use as our base dimensions
length, time and mass.  A realization of this theory is that all the matter fields in 4D
can arise from a higher-dimensional vacuum.  One starts with the vacuum Einstein equations
in 5D, and dimensionally reduces the Ricci tensor to obtain the 4D Einstein equations with
an effective stress-energy tensor \cite{wesleon}.  For this reason the theory is also called
induced-matter theory.

In this paper, we re-examine the field equations for free electromagnetic fields derived
from an action principle using a Lagrangian that is constructed from a ``generalized''
electromagnetic field tensor.  To simplify our calculations, we restrict ourselves to flat
4D spacetime.  We use the following conventions.  Upper-case Latin indices
$A,B,\ldots\in\{0,\ldots,4\}$, lower-case greek indices $\alpha,\beta,\ldots\in\{0,\ldots,3\}$
and lower-case latin indices $i,j,\ldots\in\{1,2,3\}$.  The signature of the 5D metric
is $+---\varepsilon$ with $\varepsilon\in\{-1,1\}$ employed to allow for either signature of
the extra coordinate, and the signature of the 4D spacetime metric is $+---$.  We work in
units where the speed of light $c$ and the gravitational constant $8\pi G$ are unity.

\section{Outline of the Problem}

In order to motivate our current study, we present the essential features of induced-matter
theory in a form that is useful here.  The material in this section follows the general
arguments given by Ponce de Leon in \cite{leo1}.  For specialization to diagonal 5D metrics
we refer the reader to the review \cite{low}.

The starting point of the theory is a 5D line element expressed in terms of the physical
spacetime metric $g_{\alpha\beta}=g_{\alpha\beta}(x^{\gamma},x^{4})$, electromagnetic
potentials $A_{\alpha}=A_{\alpha}(x^{\gamma},x^{4})$ and scalar field
$\Phi=\Phi(x^{\gamma},x^{4})$ in the general form
\bea
dS^{2} = \mbox{e}^{2\xi}g_{\alpha\beta}dx^{\alpha}dx^{\beta}
         + \varepsilon\Phi^{2}(dx^{4}+A_{\alpha}dx^{\alpha})^{2}.
\label{metric}
\eea
Here the scalar function $\xi=\xi(x^{\gamma},x^{4})$ is included in order to distinguish the
induced spacetime metric from the physical one.  Most often the two are equivalent, however,
the distinction is necessary in general to describe all possible physical scenarios.
Putting the 5D metric (\ref{metric}) into the 5D vacuum Einstein equations $R_{AB}=0$ we
recover ten field equations (Einstein) for the induced spacetime metric
$\mbox{e}^{2\xi}g_{\alpha\beta}$ with a source that depends on the fields
$A_{\alpha}$ and $\Phi$, in addition to $x^{4}$-derivatives of
$\mbox{e}^{2\xi}g_{\alpha\beta}$.  We also recover four field equations (Maxwell) for the
electromagnetic potentials with a source that depends on $\Phi$, and one field equation
(Klein-Gordon) for $\Phi$.

The electromagnetic potentials $A_{\alpha}$ determine the Faraday field tensor
$F_{\alpha\beta}=\partial_{\alpha}A_{\beta}-\partial_{\beta}A_{\alpha}$.  As pointed out by
Ponce de Leon in \cite{leo1,leo2}, physics in 4D should be invariant under the set of
transformations
\bea
x^{\alpha} = \tilde{x}^{\alpha}(x^{\gamma})
\quad
\mbox{and}
\quad
x^{4} = \tilde{x}^{4} + f(\tilde{x}^{\gamma})
\label{transformations}
\eea
that leave the 5D line element unchanged.  Under these transformations, however, we have that
$A_{\alpha}\rightarrow A_{\alpha}+\partial_{\alpha}f$ and
$F_{\alpha\beta}\rightarrow \hat{F}_{\alpha\beta}=F_{\alpha\beta}+(\partial_{4}A_{\beta})
(\partial_{\alpha}f)-(\partial_{4}A_{\alpha})(\partial_{\beta}f)$.  Thus for the fields to be
``gauge'' invariant we require that
\bea
\hat{F}_{\alpha\beta} = F_{\alpha\beta} + G_{\alpha\beta},
\label{generalized}
\eea
where we have defined
$G_{\alpha\beta}=A_{\beta}\partial_{4}A_{\alpha}-A_{\alpha}\partial_{4}A_{\beta}$.  This tensor
$\hat{F}_{\alpha\beta}$ is invariant under the transformations (\ref{transformations}), and
henceforth shall be referred to as the generalized electromagnetic field tensor.  If the
potentials $A_{\alpha}$ are independent of the fifth coordinate, then $G_{\alpha\beta}$ vanishes
and $\hat{F}_{\alpha\beta}\rightarrow F_{\alpha\beta}$.  In the next section, we consider free
electromagnetic fields in flat Minkowski spacetime, and derive the Maxwell equations from an
action principle using the generalized electromagnetic field tensor.  Using this framework we will
isolate an effective electromagnetic current, and we will see that the source of this current is
purely of 5D origin.

\section{Maxwell's Equations and Induced Current from Five Dimensions}

Our intention here is to study the extent to which the non-vanishing terms in
$G_{\alpha\beta}$ coming from the derivatives $\partial_{4}A_{\alpha}$ have on the 4D physics
of electromagnetism.  For this purpose we take $\xi=0$ and
$g_{\alpha\beta}=\eta_{\alpha\beta}=\mbox{diag}(+1,-1,-1,-1)$.  In order to have free
electromagnetic fields we also need to decouple the potentials from the scalar field.
Thus we set $\Phi=\mbox{constant}$.

So let us begin by taking as the Lagrangian for free electromagnetic fields without sources in
Minkowski spacetime the scalar functional
\bea
\mathcal{S} = -\frac{1}{4}\int d^{4}x\hat{F}_{\alpha\beta}\hat{F}^{\alpha\beta}.
\label{action}
\eea
(In curved 4D spacetime the above generalizes to
$\mathcal{S}=(-1/4)\smallint d^{4}x\sqrt{-g}\hat{F}_{\alpha\beta}\hat{F}^{\alpha\beta}$
where $g=\mbox{det}(g_{\alpha\beta})$.)  Here the Lagrangian density has functional form
$\mathcal{L}=\mathcal{L}[\phi_{i},\partial_{\alpha}\phi_{i},\partial_{4}\phi_{i}]$ where
$\phi_{i}$ labels the different species of fields in the theory (in this case the vector
potentials).  Variation of the action
$\mathcal{S}=\smallint d^{4}x\mathcal{L}[\phi_{i},\partial_{\alpha}\phi_{i},\partial_{4}\phi_{i}]$ gives
\bea
\delta\mathcal{S}
&=& \int d^{4}x\left[\pdiff{\mathcal{L}}{\phi_{i}}\delta\phi_{i}
                     + \pdiff{\mathcal{L}}{(\partial_{\alpha}\phi_{i})}\delta(\partial_{\alpha}\phi_{i})
                     + \pdiff{\mathcal{L}}{(\partial_{4}\phi_{i})}\delta(\partial_{4}\phi_{i})\right]\nonumber\\
&=& \int d^{4}x\left[\pdiff{\mathcal{L}}{\phi_{i}}
                     - \partial_{\alpha}\left(\pdiff{\mathcal{L}}{(\partial_{\alpha}\phi_{i})}\right)
                     - \partial_{4}\left(\pdiff{\mathcal{L}}{(\partial_{4}\phi_{i})}\right)\right]\delta\phi_{i}.
\eea
In going from the first line to the second we used
$\delta(\partial_{\alpha}\phi_{i})=\partial_{\alpha}(\delta\phi_{i})$ and
$\delta(\partial_{4}\phi_{i})=\partial_{4}(\delta\phi_{i})$, integrated by parts and discarded the
surface terms.  The principle of least action then states that the equations of motion are the critical
points of $\mathcal{S}$.  Thus from $\delta\mathcal{S}=0$ we have that
\bea
\pdiff{\mathcal{L}}{\phi_{i}}
- \partial_{\alpha}\left(\pdiff{\mathcal{L}}{(\partial_{\alpha}\phi_{i})}\right)
- \partial_{4}\left(\pdiff{\mathcal{L}}{(\partial_{4}\phi_{i})}\right) = 0.
\label{eqnsofmotion}
\eea
For the action (\ref{action}), we find through some tedious algebra that
\bea
\pdiff{\mathcal{L}}{A_{\sigma}} &=& -\hat{F}^{\sigma\alpha}\partial_{4}A_{\alpha},\\
\pdiff{\mathcal{L}}{(\partial_{\rho}A_{\sigma})} &=& -\hat{F}^{\rho\sigma},\\
\pdiff{\mathcal{L}}{(\partial_{4}A_{\sigma})} &=& -\hat{F}^{\alpha\sigma}A_{\alpha}.
\eea
The equations of motion are thus
\bea
\partial_{\alpha}\hat{F}^{\alpha\sigma} + \partial_{4}\hat{F}^{\alpha\sigma}A_{\alpha}
+ 2\hat{F}^{\alpha\sigma}\partial_{4}A_{\alpha} = 0.
\label{eqnsmotion}
\eea
If there is no dependence on the extra coordinate $x^{4}$ then the above reduce to
$\partial_{\alpha}F^{\alpha\sigma}=0$, which are the inhomogeneous Maxwell equations
for free electromagnetic fields in vacuum.  The above equations of motion are
in the form $\partial_{\alpha}\hat{F}^{\alpha\sigma}=J^{\sigma}$.  However, we began
with an action that describes free electromagnetic fields in vacuum.  Thus we
conclude that the quantity
\bea
J^{\sigma}
= \partial_{4}\hat{F}^{\sigma\alpha}A_{\alpha} + 2\hat{F}^{\sigma\alpha}\partial_{4}A_{\alpha}
\label{current}
\eea
is an effective electromagnetic current with a source that is purely 5D in origin.  This is
in line with induced-matter theory, whereby the inertial and electromagnetic properties of
matter in 4D can be explained in terms of pure geometry in a higher-dimensional world.
We note that the antisymmetry of the generalized electromagnetic field tensor implies that
$\partial_{\sigma}\partial_{\alpha}\hat{F}^{\alpha\sigma}=0$ because the second derivative
$\partial_{\sigma}\partial_{\alpha}$ is symmetric in its indices.  This means that the
effective current automatically satisfies the continuity equation
$\partial_{\sigma}J^{\sigma}=0$.  Also note that $J^{\sigma}=0$ if the cylinder condition
 (i.e. all $x^{4}$-derivatives are zero) is employed.
It is then no suprise that this current has not been previously derived in the old
Kaluza-Klein theory.

In our framework, we demand that the Lorentz force that a particle of mass $m$ and charge $q$
feels as it moves through the electromagnetic field $\hat{F}_{\alpha\beta}$ be
$\bm{F}=dp^{\alpha}/d\tau=q\hat{F}^{\alpha\beta}u_{\beta}=q(u_{0}\bm{E}+\bm{u}\times\bm{B})$
where $p^{\alpha}=m(u^{0},\bm{u})$ is the momentum four-vector, $\tau$ is proper time, $\bm{E}$
is the electric field and $\bm{B}$ is the magnetic field.  So we are naturally led to define the
electric field as $E^{i}=\hat{F}^{0i}$ and the magnetic field $B^{i}=-(1/2)\epsilon^{ijk}B_{jk}$
in terms of the field strength $B_{ij}=\hat{F}_{ij}$.  The inhomogeneous Maxwell equations can
now be written in the more familiar form
\bea
\nabla \cdot \bm{E} = \rho
\quad
\mbox{and}
\quad
\nabla \times \bm{B} - \pdiff{\bm{E}}{t} = \bm{J},
\label{imaxwell}
\eea
where we have defined the components of the effective current four-vector $J^{\sigma}=(\rho,\bm{J})$
so that
\bea
\rho &=& \partial_{4}\hat{F}^{0\alpha}A_{\alpha} + 2\hat{F}^{0\alpha}\partial_{4}A_{\alpha},\\
J^{i} &=& \partial_{4}\hat{F}^{i\alpha}A_{\alpha} + 2\hat{F}^{i\alpha}\partial_{4}A_{\alpha}.
\eea
In addition we also have the homogeneous Maxwell equations
\bea
\nabla \cdot \bm{B} = 0
\quad
\mbox{and}
\quad
\nabla \times \bm{E} + \pdiff{\bm{B}}{t} = 0
\label{hmaxwell}
\eea
that follow from the Bianchi identity $\partial_{\gamma}\hat{F}_{\alpha\beta}+
\partial_{\alpha}\hat{F}_{\beta\gamma}+\partial_{\beta}\hat{F}_{\gamma\alpha}=0$. 
Here we pause for a moment and comment on the effective current that we have derived.  From
the Bianchi identity it is evident that the magnetic field is divergence-free.  This means
that there is no magnetic charge, and hence there do not seem to exist any magnetic monopoles
in our framework.  However, Ponce de Leon notes in \cite{leo2} that under general conditions
with no restrictions on the metric $g_{AB}$ the theory does contain effective currents
corresponding to both electric and magnetic charges, thus allowing the existence of magnetic
monopoles in general.  Here we have obtained an effective current for electric charge only.
However, we have considered free electromagnetic fields in flat spacetime.  In other words,
we have neglected gravitational and dilaton interactions.  We are thus led to conclude that
the existence of magnetic monopoles depends on how the electromagnetic potentials couple to
the induced spacetime metric and scalar field.

\section{Discussion}

We would like to give some meaning to the effective current in (\ref{current}).  Here we are
mainly interested in finding a signature of this current that can be observed in a laboratory.
To this end, let us first consider an external source with current given by
$\mathcal{J}^{\sigma}=(\varrho,\bm{\mathcal{J}})$.  For this the action (\ref{action}) becomes
\bea
\mathcal{S} = -\frac{1}{4}\int d^{4}x\left(\hat{F}_{\alpha\beta}\hat{F}^{\alpha\beta}
              - A_{\alpha}\mathcal{J}^{\alpha}\right).
\eea
Then the equations of motion become
\bea
\partial_{\alpha}\hat{F}^{\alpha\sigma} = \mathcal{J}^{\sigma} + J^{\sigma}.
\label{eqnssource}
\eea
We are mainly interested in the time component of this equation.  Thus we have
$\partial_{\alpha}\hat{F}^{\alpha0}=\mathcal{J}^{0}+J^{0}$.  Now, it can be shown that
if the only non-vanishing component of the vector potential is $A_{0}=\varphi(x)$ where $x$
is a point $p\in(\mathcal{M},g_{AB})$, the time component of the induced current is
identically zero.  This follows because
\bea
\partial_{4}\hat{F}^{0\alpha}A_{\alpha}
&=& \partial_{4}\hat{F}^{00}A_{0}
    + \partial_{4}\hat{F}^{01}A_{1}
    + \partial_{4}\hat{F}^{02}A_{2}
    + \partial_{4}\hat{F}^{03}A_{3} = 0,\\
2\hat{F}^{0\alpha}\partial_{4}A_{\alpha}
&=& 2\hat{F}^{00}\partial_{4}A_{0}
    + 2\hat{F}^{01}\partial_{4}A_{1}
    + 2\hat{F}^{02}\partial_{4}A_{2}
    + 2\hat{F}^{03}\partial_{4}A_{3} = 0.
\eea
In both lines above the first term vanishes because $\hat{F}^{00}=0$, and the remaining terms
vanish because $A_{i}=0$.  Furthermore, the divergence of the generalized electromagnetic
tensor also simplifies considerably.  Here we note that
\be
\partial_{\alpha}G^{\alpha0}
&=& \partial_{0}(A^{0}\partial_{4}A^{0} - A^{0}\partial_{4}A^{0})
+ \partial_{1}(A^{0}\partial_{4}A^{1} - A^{1}\partial_{4}A^{0})\\
&\phantom{=}& + \partial_{2}(A^{0}\partial_{4}A^{2} - A^{2}\partial_{4}A^{0})
+ \partial_{3}(A^{0}\partial_{4}A^{3} - A^{3}\partial_{4}A^{0}) = 0.
\ee
So $\partial_{\alpha}\hat{F}^{\alpha0}=\partial_{\alpha}F^{\alpha0}$.  In Lorentz
gauge ($\partial_{\alpha}A^{\alpha}=0$) this becomes $\partial_{\alpha}F^{\alpha0}
=\partial_{\alpha}\partial^{\alpha}A^{0}=\partial_{\alpha}\partial^{\alpha}\varphi(x)$.
Therefore the equations (\ref{eqnssource}) are
$\partial_{\alpha}\partial^{\alpha}\varphi(x)=\mathcal{J}^{0}=\varrho$, and in the static
limit this becomes
\bea
\nabla^{2}\varphi(x) = -\varrho.
\eea
This is the Poisson equation for the potential $\varphi(x)$ of a charge distribution
with density $\varrho$.  In particular, this means that for a point charge $\varrho=q$
this is $\nabla^{2}\varphi=-q$, and therefore the Coulomb potential is unaffected by
the modifications to the Maxwell equations imposed by the extra dimension.  This is a
very interesting result, as it shows that the Coulomb potential of a point charge at
a point $p\in(\mathcal{M},g_{AB})$ remains unchanged even if the potential varies
with the extra coordinate.  This is consistent with our redefinition of the electric
field in terms of the generalized electromagnetic field tensor instead of the Faraday
field tensor.

Until now we have only dealt with classical fields.  The interactions that we observe in
nature, however, are intrinsically quantum mechanical.  So we would like to make a few
comments regarding the quantization of the electromagnetic field.  Looking at the equations
of motion $\partial_{\alpha}\hat{F}^{\alpha\sigma}=J^{\sigma}$ for free fields, this turns
out to be quite difficult due to choices of gauge.  For instance, we can employ
$\partial_{\alpha}A^{\alpha}=0$ together with $\partial_{4}A^{\alpha}=0$ so that the
equations of motion reduce to $\partial_{\alpha}\partial^{\alpha}A^{\beta}=0$.  In momentum
space (with $\partial_{\alpha}\rightarrow ip_{\alpha}$) this is
$(-p^{2}\eta_{\alpha\beta})A^{\beta}=0$ and so the photon propagator is
$-i\eta_{\alpha\beta}/p^{2}$.  This is the Feynman propagator in Lorentz gauge.  So what
if the gauge condition $\partial_{4}A^{\alpha}=0$ is not imposed?  After all, there is no
reason why this choice should be made.  Clearly this would give a somewhat complicated
expression for the photon propagator.  The derivation of the general case is beyond the scope
of the current study, and we hope to return to it in the future.  However, we wish to emphasize
that the $\partial_{4}A^{\alpha}\neq0$ could in principle be used to test for the existence of
extra dimensions in high-energy particle scattering.  Consider the following Lagrangian density:
\bea
\mathcal{L}_{QED} = -\frac{1}{4}\hat{F}_{\alpha\beta}\hat{F}^{\alpha\beta}
                    + \bar{\psi}[i\gamma^{\alpha}(\partial_{\alpha}+ieA_{\alpha})-m]\psi.
\eea
This is the action for quantum electrodynamics in Minkowski spacetime describing the
interaction of the vector potential $A_{\alpha}$ with the spinor fields $\psi$ and
$\bar{\psi}$ of mass $m$ and electric charge $|e|$.  For more details we refer the reader to
the book \cite{halmar} by Halzen and Martin.  The vector potential describes a massless spin
one gauge boson, while the spinor fields describe spin one-half fermions of opposite electric
charge and opposite helicity.  In the above Lagrangian the interaction term
$-e\bar{\psi}\gamma^{\alpha}A_{\alpha}\psi$ is unchanged, so the vertex factor remains
the same.  However, we have seen that if $\partial_{4}A_{\alpha}\neq0$ then the free Lagrangian
for the potentials is augmented by the appearance of the effective current.  The derivatives
in this current should give a modified propagator for the photon, and hence a theoretical
calculation of a scattering process may be used to isolate a signature of extra dimensions
that can be tested in the laboratory.

\section*{Acknowledgements}

This work was supported by the Natural Sciences and Engineering Research Council of Canada
(NSERC) under grant 101203.

\end{document}